\documentclass[sigconf]{acmart}

\usepackage{booktabs} 

\setcopyright{none}

\renewcommand\footnotetextcopyrightpermission[1]{}

\usepackage{flushend}

\acmConference[ ]{ }{ }{ }
\acmYear{2018}
\copyrightyear{2018}

\newcommand{\etal}{et~al.}

\begin{document}
\title{Offline Behaviors of Online Friends}

\author{Piotr Sapiezynski}
\affiliation{%
  \institution{Technical University of Denmark}
  \institution{Northeastern University}
}

\author{Arkadiusz Stopczynski}
\affiliation{%
  \institution{Technical University of Denmark}
  \institution{MIT Media Lab}
}

\author{David Kofoed Wind}
\affiliation{%
  \institution{Technical University of Denmark}
}

\author{Jure Leskovec}
\affiliation{%
  \institution{Stanford University}
}

\author{Sune Lehmann}
\affiliation{%
  \institution{Technical University of Denmark}
  \institution{Niels Bohr Institute}
}

\renewcommand{\shortauthors}{P. Sapiezynski et al}


\begin{abstract}
In this work we analyze traces of mobility and co-location among a group of nearly 1\,000 closely interacting individuals.
We attempt to reconstruct the Facebook friendship graph, Facebook interaction network, as well as call and SMS networks from longitudinal records of person-to-person offline proximity. 
We find subtle, yet observable behavioral differences between pairs of people who communicate using each of the different channels
and we show that the signal of friendship is strong enough to stand out from the noise of random and schedule-driven offline interactions between \emph{familiar strangers}.
Our study also provides an overview of methods for link inference based on offline behavior and proposes new features to improve the performance of the prediction task.
\end{abstract}

%
%

\maketitle


\section{Introduction}
The social bonds we form have a lasting impact on our lives and constitute a fundamental building block of our societies.
Social connections facilitate access to resources~\cite{granovetter1973strength}, dissemination of opinions and innovation~\cite{centola2007complex}, as well as the spread of habits~\cite{christakis2007spread,graham1991social}.
As described by Krackhardt et al.~\cite{krackhardt94}, one of the key factors in human bond formation is propinquity, physical and psychological proximity between people.
Until recently, scientists lacked tools to monitor mobility and person-to-person proximity over extended periods and at large scales.
However, with recent technological advancements such as smartphones, the Internet, and online social networks, this task has become increasingly feasible~\cite{lazer2009life}.
Data is now recorded in a more seamless fashion and with a significantly reduced involvement on behalf of the researchers.
This approach may lower the impact of the observation on the participants' behavior, compared to traditional methods.
Rather than depending on the self-reported proximity among participants~\cite{informant1979}, it is now possible to sense the interactions using sociometric badges~\cite{wu2008mining}, software running on mobile phones utilizing Bluetooth~~\cite{eagle2006reality, aharony2011social, 10.1371/journal.pone.0095978}, fixed-location sensors~\cite{Larsen2013roskilde}, mobile network towers~\cite{nsa2012} and Call Detail Records (CDRs)~\cite{dasgupta2008social}, social media check-ins~\cite{cranshaw2010bridging,scellato2011exploiting}, or geo-tagged online content~\cite{crandall2010inferring}.

In this study we show how a social network can be inferred from person-to-person proximity events tracked over time.
We use a year-long record of such interactions among a population of nearly 1\,000 students.
The events are inferred using a novel method described in Ref.~\cite{sapiezynski2017inferring}.
In lieu of self-reported relationships, we use a number of communication channels as ground-truth proxies for social ties.
Given a history of a dyad's person-to-person interactions, we are able to determine whether they are connected on Facebook or call each other, or whether they are just a pair of interacting \textit{familiar strangers}~\cite{milgram1977familiar}.
We find that the size of the group in which people meet is an important factor in accurately inferring social ties, with friends spending more time in smaller groups.
We also show that interactions between friends are less likely to follow a particular schedule, compared to interactions between non-friends.
The number of people our subjects interact with in the real world is orders of magnitude larger that the number of their online contacts.
The participants of our study attend lectures, perform group work, and socialize during events organized by the university.
Nevertheless, the signal of friendship is strong enough not to be lost in the noise of proximity interactions with strangers.

There are three central findings in this work.
First, we verify that the proxies of friendship used in computational social science (Facebook interactions, calls, and short messages) are reflected in the offline behaviors and are positively correlated with the intensity of person-to-person contact.
Second, we show that there is a surprisingly low overlap between dyads who communicate on Facebook and those who call each other, and additionally that 
offline behaviors can reveal which particular channel of communication is adopted by each dyad.
Finally, we compare the performance of a number of behavioral traits in discerning actual social ties from those imposed by class schedules.
The insights can be applied to aid research and empower social applications, but also raise important questions regarding the privacy of millions of smartphone users.

\section{Experimental design}
\subsection{The Copenhagen Networks Study}
The dataset used in this work was collected as part of the Copenhagen Networks Study (CNS)~\cite{10.1371/journal.pone.0095978}.
In CNS, we tracked mobility as well as online and offline interactions in a densely connected community of approximately 1\,000 students for two years.
All data in CNS was collected with the participants' informed consent, and with an emphasis on ensuring awareness of the complexity and sensitivity of the collected data~\cite{StopczynskiPPLL14}.
The study design --- including its security and privacy aspects, as well as the informed consent --- has been approved by the Danish Data Protection Agency.

Each participant of the study received an Android smartphone (LG Nexus 4) and installed a custom data collection app.
The collector software was based on Funf Open Sensing framework~\cite{aharony2011social} and ran in the background, regardless of a user's activity.
It collected a variety of sensor readings with high resolution, including: location estimations (every five minutes), WiFi scan results (every 15 seconds), records of phone calls and short messages, and person-to-person interactions approximated by Bluetooth visibility (every five minutes).
The data, compressed and encrypted, was periodically uploaded to a server located at the university.
In addition to data acquired directly from the participants' phones, we also gathered snapshots of participants' Facebook data every 24 hours (server-side collection), including lists of their friends, as well as likes, tags, and posts.
Further details on the study can be found in Ref.~\cite{10.1371/journal.pone.0095978}.

The high-resolution data collected in CNS offers an opportunity for unprecedented insight into the dynamics of a complex social system seen across multiple channels.
The location data allows for high-resolution fixed rate measurement of the participants' mobility, both inside the buildings and outdoors.
Calls, texts, and Facebook interactions reveal important aspects of the population's social structure.
Communication on these channels is rarely incidental, thus considering the number of interactions on these channels provides a proxy for quantifying the strength of social ties.

The starting point for this study is that discovering the internal social structure of a densely-connected population---such as the population considered here---based on physical proximity is a non-trivial task.
Non-acquainted participants in the study meet on a regular basis during lectures, in cafeterias, or at the gym.
Here, we endeavor to extract a social signal above and beyond the dense hairball of spurious interactions~\cite{sekara2016fundamental} and to reconstruct the entire social network of the participants.

\subsection{Data availability}
\begin{figure}[t!]
\center
\includegraphics[width=1\linewidth]{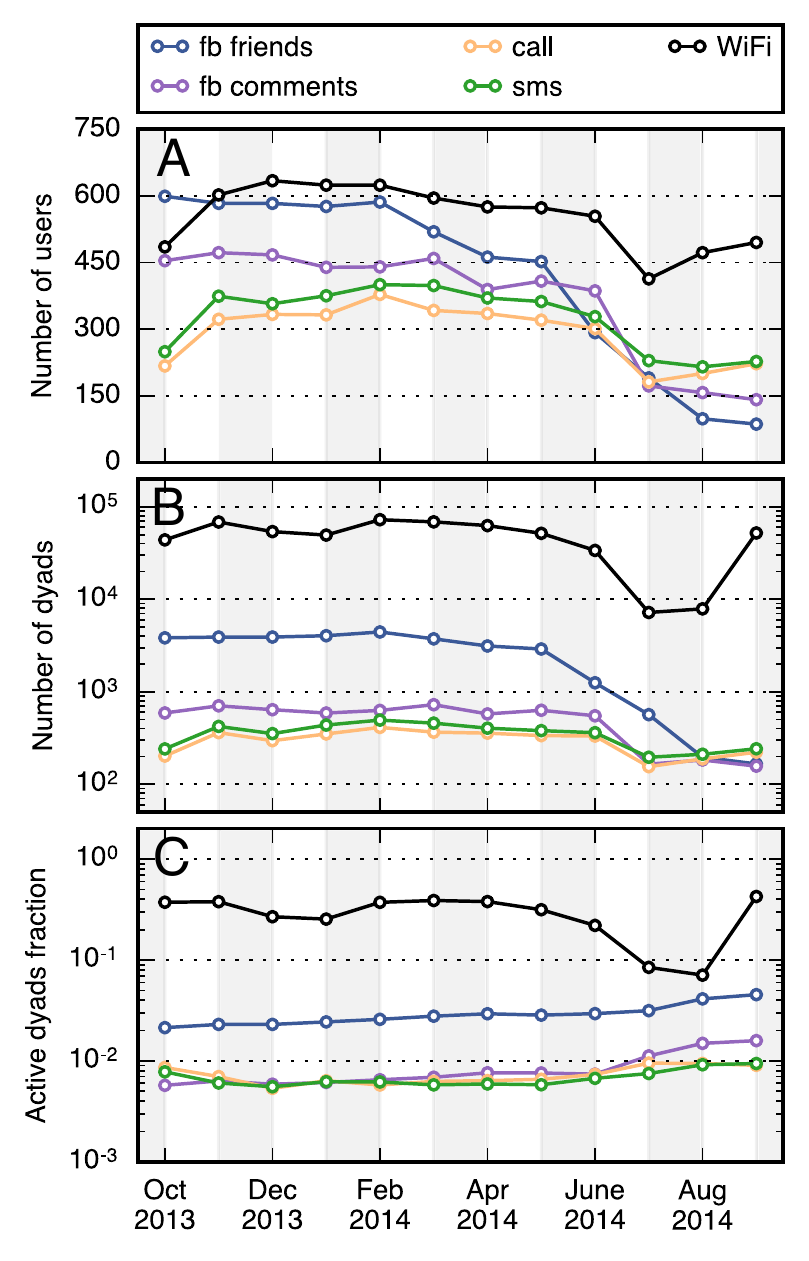}

\caption{A) The number of participants with particular sources of data as a function of time. The fact that the users did not renew the authorization to access their Facebook data causes the decline of availability. The low but stable number of people with call and sms data shows that not everybody uses the traditional communication channels. 
B) The number of dyads with interactions in each of the data sources. This shows the severity of the class imbalance problem: in the peak month of February 2014 there are 72570 dyads interacting in physical space (as inferred from WiFi) but only 412 only that exchanged phone calls. 
C) The fraction of dyads who interacted among all possible dyads. Because all participants in the study are students at the same university, as many as 37\% of possible dyads actually interact in the offline world. At the same time only 0.6\% of dyads between people who use the call functionality call each other. Note that while the fraction of active person-to-person dyads drops in the vacation period (July-August 2014), such behavior is not seen in the other communication channels. 
}
\label{fig:numberofusers}
\end{figure}

\begin{figure}[t!]
\center
\includegraphics[width=1\linewidth]{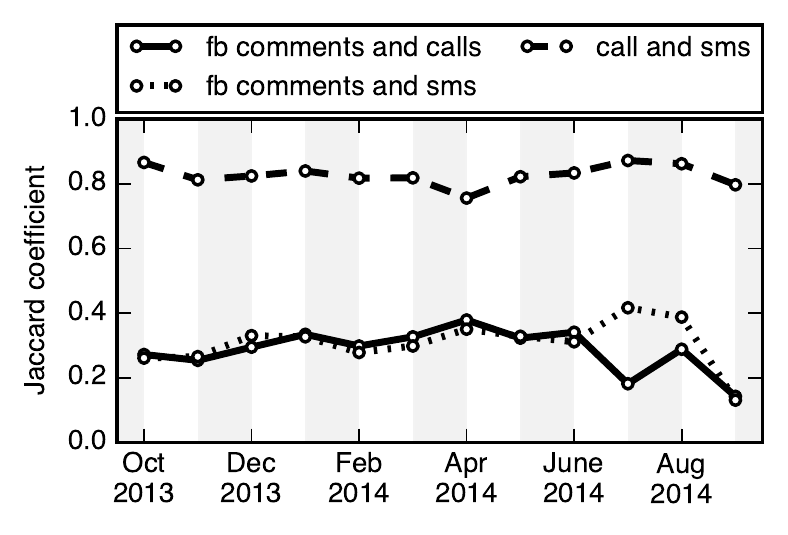}

\caption{There is a similar number of dyads that call, send messages, and exchange comments on Facebook. Among dyads formed by users who use three networks there is a much higher overlap between calling and messaging dyads, than dyads who interact on Facebook and call or message.}
\label{fig:network_comparison}
\end{figure}

Figure~\ref{fig:numberofusers} summarizes the properties of the dataset by showing the number of users (A), dyads (B), and the density of all five networks (C): person-to-person, Facebook friendship graph, Facebook interaction graph, sms, and call.
Note in Figure~\ref{fig:numberofusers}A that behavioral (WiFi) data is available for a significantly larger number of users than the ground truth data.
The number of users that provided Facebook data declines over time, as some failed to renew the authorization every three months, as required by the Facebook API.
The relatively low number of users with calls and text messages is not caused by the users failing to collect this data: it reflects that these forms of communication are less popular among the experiment participants.
Figure~\ref{fig:numberofusers}B emphasizes that the set-size of interactions in different communication channels are highly imbalanced: there are an order of magnitude more links in the Facebook friendship network compared to call~/~sms and the Facebook interaction network, and an order of magnitude more interactions person-to-person compared to Facebook friend links.
As shown in Figure~\ref{fig:numberofusers}C, contacts via communication networks are driven by the academic schedule to a lesser extent than person-to-person interactions: the fraction of active dyads does not decrease during the summer time, in contrast to the person-to-person network.
Finally, Figure~\ref{fig:network_comparison} shows that there is a surprisingly small overlap between the dyads who interact on Facebook and those who call or message each other, and that difference persists over time.
Both telecommunication exchanges and Facebook comments have been used as proxies for friendship in scientific literature.
This low overlap indicates that relying on only one of these channels is likely to yield a distorted picture of the underlying friendship network.

\subsection{Proximity inference}
In this work, we use proximity events inferred from WiFi data.
The details of the inference procedure can be found in Ref.~\cite{sapiezynski2016inferring}.
The findings presented here do not depend on using WiFi as a proxy for person-to-person proximity and alternative methods, such as Bluetooth or RFID sensing, would work equally well.
However, when using these alternative methods for social sensing, however, an additional step is necessary in order to determine the location of each interaction.


\section{Methods}
The task is to infer social ties in a group of people, given their history of interactions in physical space.
In lieu of self-reported relationships, we infer four proxies of social ties: Facebook friendships, Facebook interactions (comments on each other's content), phone calls, and text messages.
For simplicity we treat all networks as non-directional and assume that the links are reciprocated (although we acknowledge that in real-world networks the perceived friendships are not always reciprocal~\cite{10.1371/journal.pone.0151588}).
To accomplish the task, we take the following steps:
\begin{enumerate}
\setlength{\itemsep}{1pt}
\setlength{\parskip}{2pt}
\setlength{\parsep}{2pt}
\item We contextualize each meeting by describing its social makeup, timing, and location.
\item We create a set of features that describe and summarize the properties of the person-to-person interactions of each dyad among participants.
\item We train supervised machine learning models on a subset of dyads and infer the links among the remainder of population. We do so separately for each of the four types of links.
\end{enumerate}

We infer the links for each month from the interactions during the same month. 
Additionally, we answer the following questions: 
(1) does knowing the history of offline interactions for more than one month increase performance of the inference?
(2) can we infer (using the history of offline interactions) which communication tool a dyad uses, knowing that they do communicate either through Facebook comments or calls (but not both)?

\subsection{Features}
For each dyad $A,B$ we define 16 features grouped into the following categories (see Table~\ref{tab:social_features} for an overview):

\paragraph{Time spent together in various contexts.}
After reducing the time resolution to one minute, we compute \textbf{total time together} in minutes for dyad.
Using the presence or lack of routers with the network name (SSID) of {\tt dtu}, which is the university network, we describe the location of each meeting and calculate the time \textbf{on campus} and \textbf{outside of campus}.
Additionally, assuming that the access point observed the most is one's home router, we compute the time the two people spent together \textbf{at home} of one of the individuals.
Following the intuition that friends meet in smaller groups~\cite{wang2011human}, we weigh each meeting by the size of the union of people $A$ and $B$ met within 300 seconds of the meeting and then sum the co-occurrences \textbf{weighted by the number of people} present.
In a similar way, we weigh each meeting by the number of access points as a proxy of a population density~\cite{10.1371/journal.pone.0130824} and then sum the co-occurrences \textbf{weighted by the number of APs}.

\begin{table}[t]
  \centering
  \begin{tabular}{p{30mm}p{40mm}}
  	\toprule
  	category & features \\
  	\toprule
  	\textbf{Time spent together} & total time together, on campus, outside of campus, at home, weighted by the number of people, weighted by the number of APs \\ \midrule
  	\textbf{Regularity} & entropy of hour of the day, entropy of the day of the week, entropy of the hour of the week, mean time between meetings, median time between meetings, entropy of locations\\ \midrule
  	\textbf{Network similarity} & overlap among top 5 contacts, overlap among top 15 contacts, overlap among top 25 contacts, overlap among top 50 contacts \\  \midrule

  \end{tabular}
  \caption{Overview of the features used to infer social ties. Each feature has three variants: total, in-role (considering only interactions during working hours on weekdays), and extra-role (considering only interactions outside of working hours and on weekends).}
  \label{tab:social_features}
\end{table}

\paragraph{Regularity of meetings.}
In a university setting, many people meet during classes without necessarily forming social ties.
To distinguish between `organic' and on-schedule meetings, we calculate Shannon's \textbf{entropy of hour of the day}, \textbf{entropy of the day of the week}, \textbf{entropy of the hour of the week} of each dyads' meetings.
Dyads who meet only on particular days and during work hours have a low entropy, whereas dyads meeting irregularly and at unexpected times have higher entropy scores.
Additionally, we compute the \textbf{mean-} and \textbf{median time between meetings} of each dyad: we expect friends to meet more often, even if for a short time, and non-friends to meet rarely, but possibly for longer periods at a time (e.g. at classes lasting up to 4 hours each).
Furthermore, we report the \textbf{entropy of locations} in which the meetings take place; 
intuitively, if two people always meet in the same location they are less likely to be friends than if they are seen together in different places.
Using entropy of time and location of meetings have been previously described as the Entropy Based Model~\cite{Pham2013}.

\paragraph{Network similarity.}
It is commonly assumed that social relations often are transitive: if $A$ is friends with $C$ and $C$ is friends with $B$, then $A$ is likely to be friends with $B$.
We therefore measure the Jaccard similarity of top contacts between $A$ and $B$, assuming that the more similar their top contacts are, the more likely $A$ and $B$ are to be friends.
We use the values of \textbf{overlap among top 5 contacts}, \textbf{overlap among top 15 contacts}, \textbf{overlap among top 25 contacts}, and \textbf{overlap among top 50 contacts} (noting that extending the search beyond 50 top contacts does not increase the performance of the models).
The neighborhood similarity has been previously exploited in the problem of link prediction for example in~\cite{cranshaw2010bridging}.

Each of the 16 features has three variants: total, in-role (considering only interactions during working hours on weekdays), and extra-role (considering only interactions outside of working hours and on weekends).
Previous research showed that this distinction is crucial and that extra-role interactions are more indicative of friendship~\cite{eagle2009inferring}.
We train a Random Forest classifier for each month of the data using the four networks (Facebook friendship, Facebook interactions, call, sms) as ground truth.
We have verified that other classifiers (including Gradient Boosting and thresholding on Logistic Regression) yield similar results.
Here, we focus more on identifying the behaviors indicative of friendship than selecting the best performing classifier.

\subsection{Inference performance}
We quantify the performance of our classifiers using two popular measures: Area Under Receiving Operator Curve ($AU ROC$) and Matthew's Correlation Coefficient ($MCC$).
$AU ROC$ can be interpreted as follows: given a dyad of friends and a dyad of non-friends, what fraction of times does the classifier rank the friends higher?
This metric provides an idea of how different a typical dyad of friends is from a typical dyad of non-friends.
However, because of the severe class imbalance problem (see Figure~\ref{fig:numberofusers}C) a large number of non-friends can still be misclassified as friends.
Therefore, we also provide the $MCC$ measure (Matthew's Correlation Coefficient), defined as follows:
\begin{equation}
MCC = \frac{TP\cdot TN - FP\cdot FN}{\sqrt{(TP+FP)(TP+FN)(TN+FP)(TN+FN)}},
\end{equation}
where $TP$ (true positives) is the number of friends correctly identified by the classifier, $TN$ (true negatives) is the number of correctly identified non-friends, $FP$ (false positives) is the number of non-friends incorrectly classified as friends, and $FN$ (false negatives) is the number of friends that the classifier missed.
The $MCC$ does not have a straightforward interpretation similar to $AU ROC$.
However, its value is influenced by the raw count of false positives and thus reflects the problem of class imbalance.
A value of $MCC$ close to 0 indicates that a classifier should not be used, while a value closer to 1 shows that the classifier overcomes the imbalance problem.


\section{Results}
This section is divided into four parts.
First, we quantify how likely a dyad connected through each ground truth social network is to interact in the physical space.
Second, we infer links in each of the reference networks from a number of features pertaining to offline person-to-person interactions.
Third, we show that basing the inference on more than one month of data increases the performance.
Finally, we show that the observable behaviors can be used as a base for discerning dyads who interact on Facebook from those who call.

\subsection{Offline interactions of online friends}
We find that all four networks---Facebook friendships, Facebook interactions, sms, and call---can be reliably inferred from the close proximity events collected over a year.
Our data suggests that there are different levels of psychological propinquity necessary for an edge to exist in each of the four networks.
The population we study is better connected on Facebook than via sms and calls, which indicates a lower threshold for becoming \emph{friends} online than maintaining telecommunication exchanges.

\begin{figure}[t!]
\center
\includegraphics[width=1\linewidth]{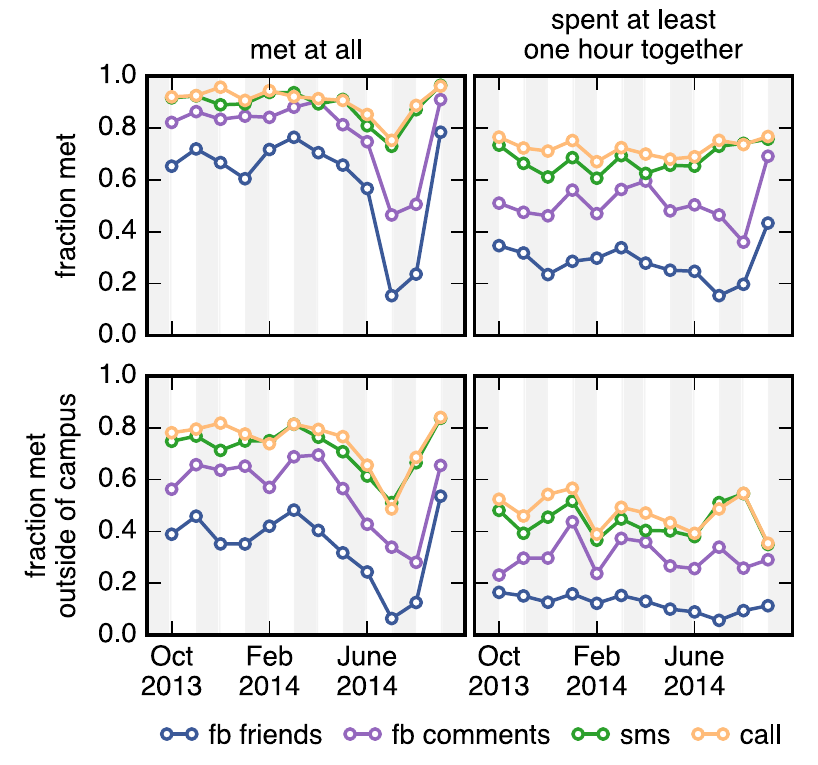}

\caption{A vast majority of people who contact each other via phone/sms also meet in the real world, often outside of campus and for longer periods. People who interact on Facebook do as well, but only a smaller subset of them.}
\label{fig:met}
\end{figure}

Furthermore, we find that dyads who actively communicate (rather than just being friends on Facebook) also interact in the offline world with higher intensity.
For example in March 2014:
\begin{itemize}
\setlength{\itemsep}{1pt}
\setlength{\parskip}{2pt}
\setlength{\parsep}{2pt}
\item \textbf{Facebook friends. } 73\% of Facebook friends met at least once during the month, including 43\% who met outside of campus; 35\% spent at least one hour together.
\item \textbf{Facebook friends, who actively interact online. } 89\% of Facebook friends who interact on Facebook met at least once during the month, including 64\% who met outside of campus; 58\% spent at least one hour together. 
\item \textbf{SMS friends. }94\% of sms contacts met at least once, including 80\% who met outside of campus; 68\% spent at least one hour together.
\item \textbf{Call friends. }93\% of call contacts met at least once, including 80\% who met outside of campus; 71\% spent at least one hour together.
\end{itemize}
The results for each month are presented in Figure~\ref{fig:met}, revealing that the tendency holds during other months as well, with higher fraction of telecommunication contacts meeting on and outside of campus, and spending more time together.

\begin{figure*}[t!]
\center
\includegraphics[width=1\linewidth]{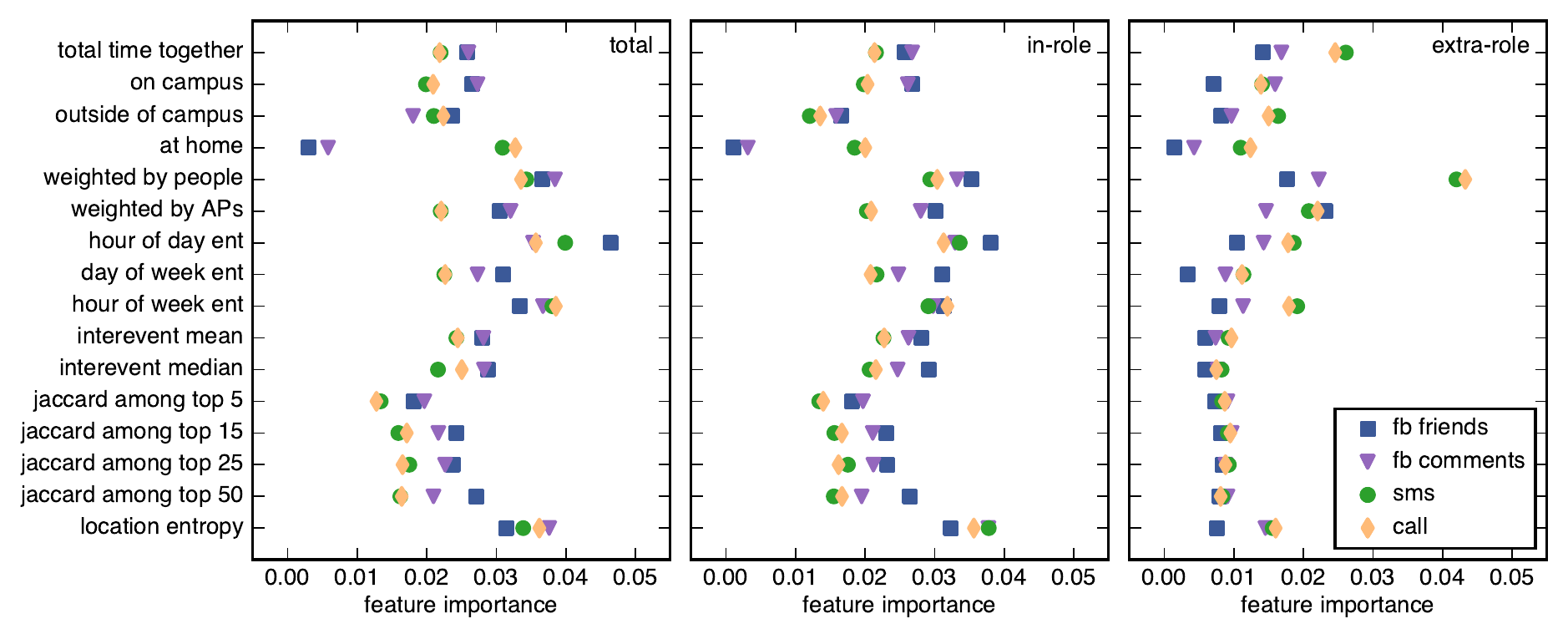}

\caption{Relative importance of features in predicting the three kinds of links. The most important feature for predicting call/sms networks it is the extra-role time weighted by the number of people present. In-role interactions are more important for inferring Facebook links than extra-role. Time at home is consistently the least important feature for inferring Facebook ties. The presented values are median importances of 10 runs of five-fold cross validation training of a Random Forest Classifier.}
\label{fig:importances_friendships}
\end{figure*}

\subsection{Inferring social networks from offline interactions}
\begin{figure}[t!]
\center
\includegraphics[width=1\linewidth]{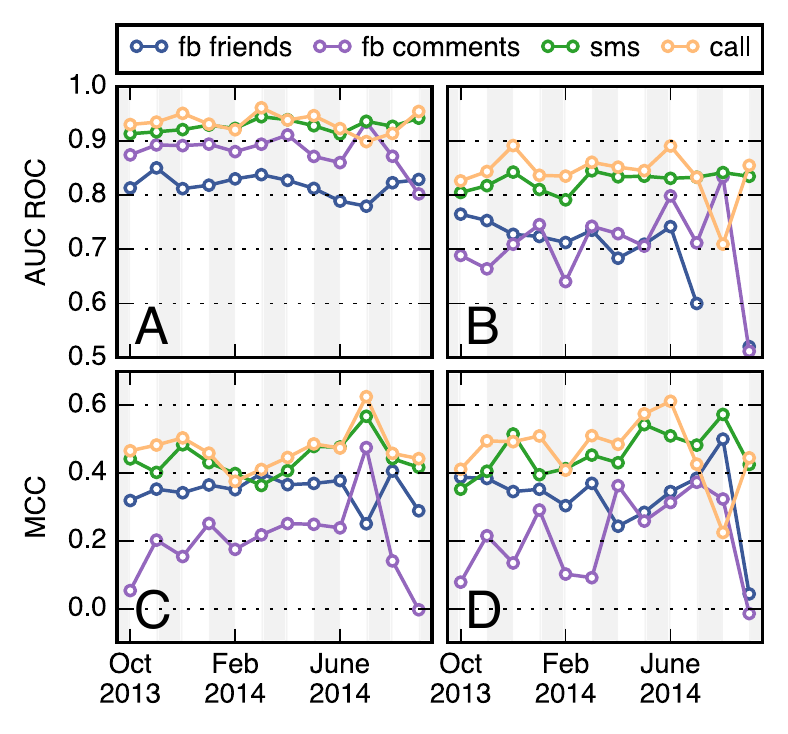}
\caption{Results of inferring friendship networks among all students (A and C) and students who study the same major (B and D). Evidently, it is possible to infer friendships even among people who, because of the class schedule, spend multiple hours per day together.}
\label{fig:roc_mcc}
\end{figure}

Given the 16 features, we infer the four kinds of friendships among the study participants. 
We perform a five-fold cross validation procedure with a random forests classifier.
In Figure~\ref{fig:roc_mcc}A we report the mean $AU ROC$ and $MCC$ scores (see Methods for interpretation) for each of the models in each prediction task.
Additionally, we perform a similar procedure, this time only investigating links among people studying the same majors.
The results reported in Figure~\ref{fig:roc_mcc}B and \ref{fig:roc_mcc}D indicate that there is a strong signal of friendship even among non-friends who spend multiple hours per day together due to class schedules.
AU ROC scores indicate that the difference between a pair Facebook typical friends and a pair of typical non-friends is less pronounced in the behavioral data that it is the case with interaction (Facebook comments, call, sms) networks, see Figure~\ref{fig:roc_mcc}.
This finding is consistent with previous research on strength of Facebook ties~\cite{wilson2009user,jones2013inferring}.
Finally, we present the importance of each feature as estimated by the Random Forest Classifier in Figure~\ref{fig:importances_friendships}.
The reported importances allow us to compare the behaviors indicative of friendship approximated via the different channels.
The time spent together in small groups in extra-role contexts is by far the most important feature for inferring the call and sms networks, but not for Facebook networks.
Furthermore, we note that home visits have much lower importance for inferring the Facebook networks than for call and sms.
In general terms, Figure~\ref{fig:importances_friendships} indicates that the extra-role features are important for inferring sms and call networks while Facebook networks (both declared and communication) are better predicted using in-role features.

\begin{figure*}[t!]
\center
\includegraphics[width=1\linewidth]{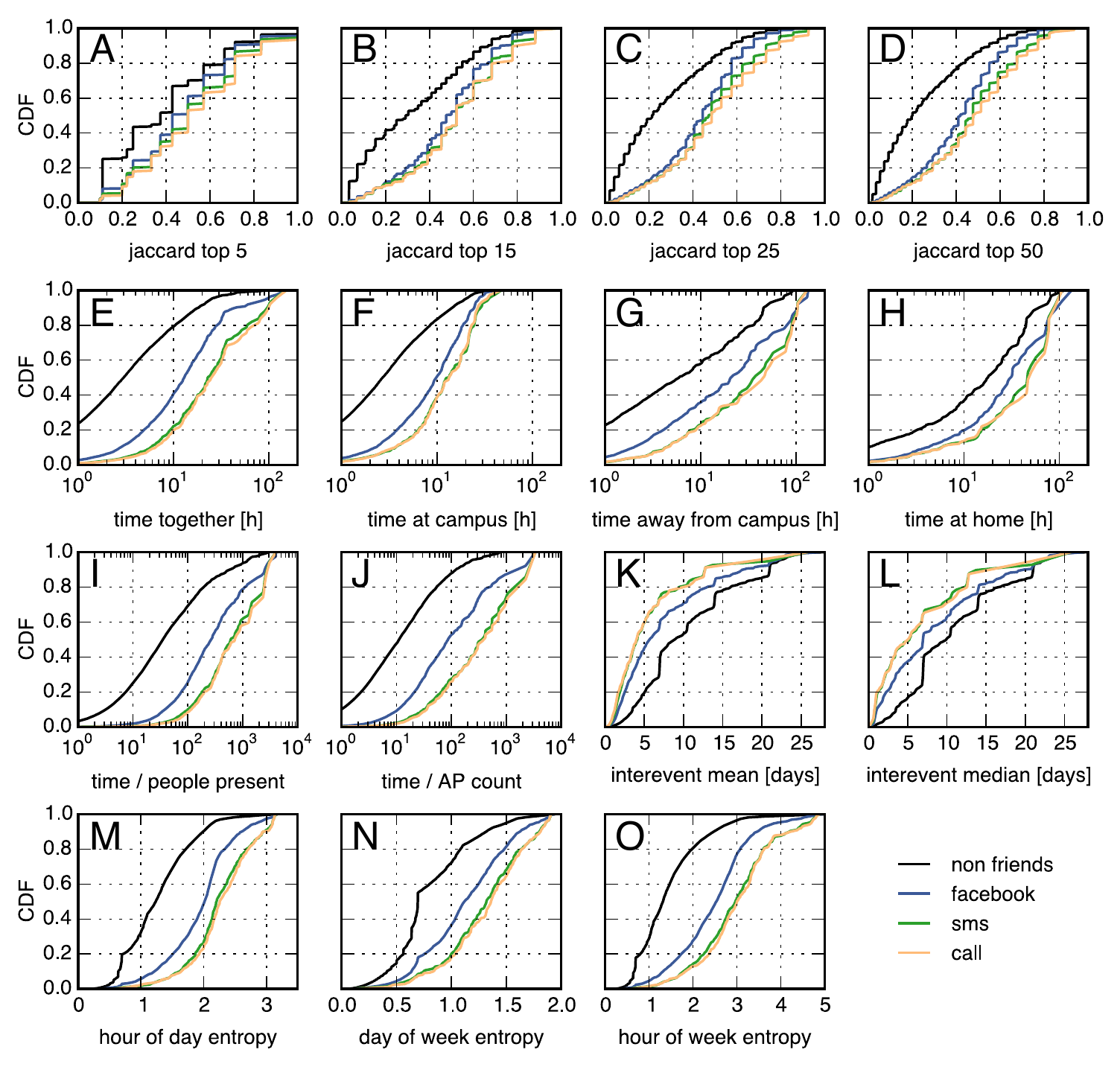}
\caption{Compared to Facebook friends (blue line) and Facebook non-friends (black line) people who call (yellow line) or message (green line) each other spend more time together (E--H) especially with only few others around (I, J), meet more often (K, L), and irregularly (M--O). They meet similar people (A--D). Behavioral signatures of Facebook friends lie between those of calling friends and people who are not friends on Facebook. The ties of people who call each other appear slightly more pronounced in face to face meetings than those who message each other.}
\label{fig:all_features}
\end{figure*}

To better understand the driving forces behind each kind of friendship we investigated, we plot the cumulative distribution of values of each feature for the different relationships in Figure~\ref{fig:all_features}.
We find that across all aspects, people who call or text each other (yellow and green lines) express a stronger tie in physical space than other types of friends, and non-friends (black line).
Figure~\ref{fig:all_features} shows that who call or text tend to spend most time together (E), both at campus (F) and outside (G), and at each others' homes (H).
We also notice that they tend to meet with fewer people present (I) and in places with a lower population density (J).
Their inter-event time is lowest among the four groups (K, L), which means that they not only spend more time together, but also meet more frequently.
Higher entropy values (M--O) indicate that the timing of their meetings has less of a scheduled character than it is the case with other relationships.
We also observe that call friends also have most similar friends.
Behavioral signatures of Facebook friends (blue line) are somewhere between those of calling friends and people who are not friends on Facebook.

\subsection{Long-term relationships}
\begin{figure}[t!]
\center
\includegraphics[width=1\linewidth]{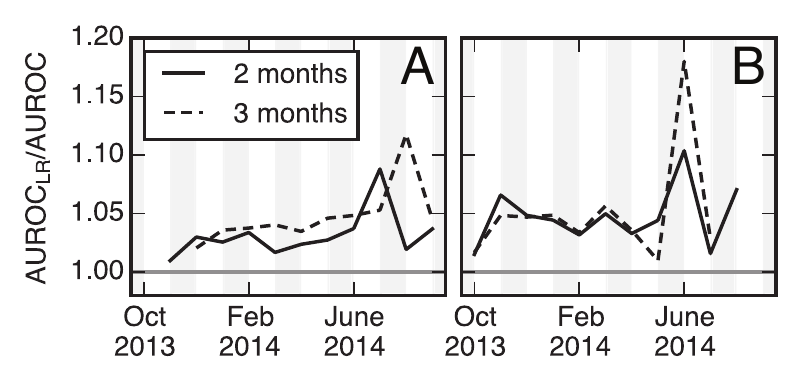}
\caption{Adding information from previous (A) and future (B) months increases the performance of predicting links (here: calls) throughout the year.}
\label{fig:log_reg}
\end{figure}

Finally, we investigate whether extending the observation time helps in the inference task. 
We build a matrix in which each row corresponds to a dyad and each column represents a month.
For each month we train a random forest classifier to infer links in the call network.
The classifier estimates and reports the probability of the existence of each link.
We store this probability in the corresponding cells of our matrix.
Then, for each month, we build a Logistic Regression model which uses these probabilities for the month and $N$ months before (Figure~\ref{fig:log_reg}A) or after (Figure~\ref{fig:log_reg}B).
Figure~\ref{fig:log_reg} reports the increase in area under receiver operator curve introduced by exploiting more than one month of data.
Using past or future data increases the performance of the inference for all months.

\subsection{Facebook friends vs. call friends}
We have already indicated that importance of features in the inference task depends on the friendship definition.
In this section we ask the following question: 
knowing the history of offline interactions of two people who communicate, can we determine which channel of communication they choose?
In order to answer this question, in each month of data we find dyads which communicate using one channel but not another.
We then build a model from the same 48 features we used for friendship inference and classify the type of friendship.
We find that it is possible, to a certain degree, to discern people who interact on Facebook from people who call and send messages.
However, possibly due to the very high overlap between dyads who call and send messages (see Figure~\ref{fig:network_comparison}) our model fails to tell these two groups apart. 
We present the results in Table~\ref{tab:friendship_types}.
We refrain from reporting the importance of each feature in the three tasks for brevity.
We note, however, that the total time weighted by the number of people was the most important feature in determining the preferred channel of communication (Facebook vs. call or sms).

\begin{table}[t]
  \centering
  \begin{tabular}{r c c}
    \toprule
  	comparison & dyads count & AUC ROC \\
  	\toprule
  	\textbf{fb comments vs. calls} & 2019 & 0.71\\
  	\textbf{fb comments vs. sms} & 2198 & 0.66\\
  	\textbf{calls vs. sms} & 1025 & 0.55\\
  	\midrule
  \end{tabular}
  \caption{The observable behaviors allow us to partially recognize friendship types: the behaviors of fb-only dyads are measurably different from behaviors of calling dyads. }
  \label{tab:friendship_types}
\end{table}

\section{Related work}

\paragraph{Inferring social ties from co-presence events.}
The work presented here connects to an existing literature studying physical co-presence. 
In their seminal work~\cite{eagle2009inferring}, Eagle~\etal, were the first to explore the relationship between self-reported social ties and behavioral data collected through smartphones.
Their analysis revealed a stronger correlation between the reported friendships and \textit{extra-role} (off campus, off hours) than \textit{in-role} meetings ($\rho=0.35$ and $\rho=0.08$ respectively).
Crawnshaw~\etal\ extended the inference approach to include context beyond the simple on/off campus indication~\cite{cranshaw2010bridging}.
The latter model included popularity of interaction locations, temporal entropy of the meetings, and neighborhood similarity between the nodes, and it outperformed the approach of Eagle~\etal.
Further, Wang~\etal~\cite{wang2011human} have shown that even co-locations inferred from comparatively low-resolution CDR data can be used to infer social ties.

In parallel to these developments, researchers have also worked on the link prediction problem in settings where the continuous behavioral data is unavailable.
Crandall~\etal\ investigated the relationship between the number of unique locations visited by two people and the probability of them being friends in a photo-sharing service~\cite{crandall2010inferring}.
Scellato~\etal\ extended this approach by introducing additional, inferred properties of locations shared among two people, such as the social entropy~\cite{scellato2011exploiting}.
Other works have shown that the probability of friendship decreases with growing geographical distance~\cite{liben2005geographic}, that clusters of friends tend to live nearby~\cite{scellato2010distance}, and that friends meet in diverse locations~\cite{Pham2013}.
There have also been developments into coupling the social and the mobility data beyond the task of link prediction.
Intuitively, since maintaining a bond requires physical proximity, some of people's mobility is driven by social factors.
Several works argue that many non-routine travels observed in real data can be attributed to individuals seeking interaction with their social contacts~\cite{grabowicz2014entangling,toole2015coupling,cho2011friendship}.

\paragraph{Communication networks as proxies for real-world relationships. }
In this article we rely on networks of Facebook friendship and interactions as well as telecommunication networks as proxies for the existence of social ties.
Below we discuss existing literature describing the applicability of such data in this context.

Wiese~\etal~\cite{wiese2015you} compared phone networks and self-reported friendships of 40 subjects.
They found that while frequent communication indicates strong ties, lack of communication does not necessarily indicate a weak tie. 
Among other contributing factors they list the realization that people use multiple channels of communication (including face-to-face meetings) and their phone networks do not fully describe their social networks.
This finding might imply that many dyads who our models misclassifies as `call friends', are in fact friends, but use different communication means.

While studying users of Facebook, Golder~\etal~\cite{golder2007rhythms} and Wilson~\etal~\cite{wilson2009user} found that only a fraction of links present in the social graph represent dyads which interact actively on Facebook. 
Wilson recommends using the interaction graph instead of the declared friendships to better model the underlying social network.
These insights were further confirmed by Jones' research on inferring self-reported friendship ties from online interaction data~\cite{jones2013inferring}.
They found that the strength of tie is correlated with the intensity of contact on Facebook, especially with commenting each other's wall content.
Furthermore, Jones found that private messages, to which we do not have access in this study, do not constitute a better indicator of real word friendship than wall posts.

Given the research described in this section, we believe that the existence of phone communication links can be treated as friendship signal among our population.
We further confirm the findings from Wilson and Jones showing that Facebook interaction networks are more predictable from offline behavioral data than the Facebook friendship graph.

\section{Limitations}
Our study does not provide access to self-reported relationships, we rely on communication channels for ground-truth.
It would be an oversimplification to claim that two people calling one another or commenting on each other's content online are necessarily friends.
Moreover, some friends might choose to communicate through means we have no access to (email, instant messaging, etc.).
Our ``ground-truth'' unfortunately does not contain such cases, and our models perform worse as a result.
There is, however, a body of research indicating that friendship does, in fact, manifest itself through the communication channels which we considered~\cite{eagle2009inferring,jones2013inferring,wiese2015you}.
We believe that going towards large-scale experiments, scientist will increasingly need to rely on observable behaviors as proxies for cognitive relationships.

\section{Conclusion}
In this work we explored the interplay between offline and online interactions of a tightly-knit group of nearly 1\,000 students.
We found that there are two orders of magnitude more links in the personal proximity network than those expressed through Facebook comments or phone calls.
Despite this imbalance, it is possible to identify the signal of friendship within this pool of interactions.
It has previously been shown that both Facebook interactions and phone calls reflect the existence of social ties.
In this work we show that these reflections are not equivalent: dyads in our dataset tend to choose one of the two means of communication.
Furthermore, we found that this choice is not random and can be predicted from offline interaction data.
The relationships social beings form are instrumental to our well-being, provide access to resources, and influence our chances of success in life.
This work shows that these ties can be inferred at scale from observable data on offline interactions.

\section*{Acknowledgements}
The authors would like to thank Aniko Hannak and Jana Huisman for their feedback on the manuscript.
In this work we used the machine learning models implementations from scikit-learn Python package~\cite{scikitlearn}.
This work was supported by the Villum Foundation [Young Investigator Program ``High Resolution Networks'' grant (to S.L.)], The Danish Council for Independent Research [Sapere Aude Program ``Microdynamics of Social Systems'' (to S.L.)], and the University of Copenhagen (UCPH Excellence Programme for Interdisciplinary Research ``Social Fabric'' grant).

\bibliographystyle{ACM-Reference-Format}
\bibliography{bibliography}

\end{document}